\documentclass[twocolumn,amsmath,amssymb,nofootinbib,eqsecnum,tighten,prb]{revtex4-1}

\usepackage{graphicx}
\usepackage{dcolumn} 
\usepackage{bm, bbm}      
\usepackage{curves}
\usepackage{epic}
\usepackage{wasysym}
\usepackage{epsf, times}
\usepackage{subfigure}
\usepackage{amsthm}

\def\openone{\leavevmode\hbox{\small1\kern-4.2pt\normalsize1}}

\def\DTO{Dy$_2$Ti$_2$O$_7$}
\def\HTO{Ho$_2$Ti$_2$O$_7$}
\def\DeltaMC{\Delta_{\rm MC}}

\newcommand{\beq}{\begin{equation}}
\newcommand{\eeq}{\end{equation}}
\newcommand{\bea}{\begin{eqnarray}}
\newcommand{\eea}{\end{eqnarray}}
\newcommand{\bfig}{\begin{figure}}
\newcommand{\efig}{\end{figure}}

\begin{document}


\title{
Debye-H\"{u}ckel theory for spin ice at low temperature
      }

\author{
C. Castelnovo$^{1,2}$
}
\author{
R. Moessner$^3$
       }
\author{
S. L. Sondhi$^4$
       }
\affiliation{
$^1$
Rudolf Peierls Centre for Theoretical Physics, University of Oxford,
Oxford, OX1 3NP, United Kingdom
}
\affiliation{
$^2$
SEPnet and Hubbard Theory Consortium, Department of Physics,
Royal Holloway University of London,
Egham TW20 0EX, United Kingdom}
\affiliation{
$^3$
Max-Planck-Institut f\"ur Physik komplexer Systeme,
Dresden, 01187, Germany
            }
\affiliation{
$^4$
Department of Physics, Princeton University,
Princeton, NJ 08544, USA
            }

\date{\today}

\begin{abstract}
At low temperatures, spin ice is populated by a finite density of magnetic 
monopoles---pointlike topological defects with a mutual magnetic Coulomb 
interaction. We discuss the properties of the resulting magnetic Coulomb 
liquid in the framework of Debye H\"{u}ckel theory, for which we provide a 
detailed context-specific account. 
We discuss both thermodynamical and dynamical signatures, and compare 
Debye H\"{u}ckel theory to experiment as well as numerics, including data for 
specific heat and AC susceptibility. We also evaluate the entropic Coulomb 
interaction which is present in addition to the magnetic one and show that it 
is quantitatively unimportant in the current compounds. 
Finally, we address the role of bound monopole anti-monopole pairs and derive an expression for the monopole mobility. 
\end{abstract}

\maketitle
%
%

\section{\label{sec: intro}
Introduction
        }
Spin systems with long-range interactions, where each spin interacts with all
others, present a formidable challenge to theoretical analysis. While
simplifications occur in the limit of infinite range interactions, the case of
dipolar interactions in three spatial dimensions is particularly complex due
to their (non-integrable) algebraic decay combined with angular dependence on
the spin direction~\cite{Ewald_refs}. As the determination of the behaviour of
even a spin model with only short ranged competing interactions can pose
a non-trivial problem, it is a priori not obvious how long-range
interactions can be treated.

\begin{figure}[ht]
\begin{center}
\includegraphics[width=0.99\columnwidth]
                {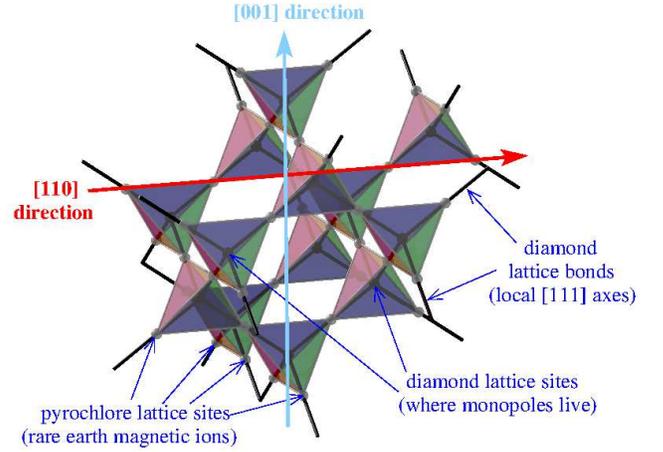}
\end{center}
\caption{
\label{fig: lattice details}
The magnetic moments in spin ice reside on the sites of the pyrochlore 
lattice, which consists of corner sharing tetrahedra. 
These sites are at the same time the midpoints of the bonds of the diamond 
lattice (black) defined by the centres of the tetrahedra. 
The Ising axes are the local [111] directions, which point along the 
respective diamond lattice bonds. 
The bonds of the pyrochlore lattice are in the [110] directions, while a 
line joining the two midpoints of opposite bonds on the same tetrahedron 
defines a [100] direction. 
}
\end{figure}

A remarkable counterexample to this case for pessimism is provided by spin 
ice~\cite{Bramwell2001}, a dipolar Ising magnet on the pyrochlore lattice 
that fails to order down to the lowest temperatures accessed. 
To a fine approximation, which we detail below, spin ice is governed by a 
model dipolar Hamiltonian about which quite a lot is known, 
\bea
H
&=&
J_{\rm ex}^{\rm nn} 
\sum_{\langle i j \rangle}
  {\bf S}_i \cdot {\bf S}_j 
\nonumber \\ 
&+&
\frac{\mu_0}{4 \pi} \sum_{i < j}
\left[
  \frac{{\bf S}_i \cdot {\bf S}_j}{ r_{ij}^3}
  -
  \frac{3({\bf S}_i \cdot {\bf r}_{ij})({\bf S}_j \cdot {\bf r}_{ij})}
       {r_{ij}^5}
\right]
\,,
\nonumber
\eea
where $J_{\rm ex}^{\rm nn}$ is the exchange interaction truncated at the 
nearest-neighbour level, the spins ${\bf S}_i$ point parallel to the local 
[111] axis (see Fig.~\ref{fig: lattice details}), and $\mu_0$ is the vacuum 
permeability. 
The rare earth spins ${\bf S}_i$ have typically a dipole moment of 
approximately $10~\mu_B$ ($\mu_B$ = Bohr magneton). 

Most prominently, the model Hamiltonian has an extensive set of ground states 
which can be specified by a purely local ``ice rule''. Their entropy is 
known to an excellent approximation due to Pauling's work already in the 
context of water ice and it has been observed 
experimentally~\cite{Ramirez1999}. 
The $T \rightarrow 0$ static correlations are averages over this ground state 
manifold and their long distance forms are known as they are described 
by an emergent gauge field in the Coulomb 
phase~\cite{Huse2003,Moessner2003b,Isakov2004,Hermele2004,Henley2005}, 
which have also been observed 
experimentally~\cite{Fennell2009,Morris2009,Kadowaki2009}.

At low temperatures the physics of the system turns out to allow a further 
simplification. The excitations about the ground state manifold take the form 
of magnetic monopoles---pointlike defects that interact via a magnetic 
Coulomb interaction energy which is independent of the background spin 
state~\cite{Castelnovo2008}. 
In this regime, the magnetic monopoles are sparse, as their number is 
suppressed on account of their excitation gap. This in turn has two 
implications. Firstly, the static correlators continue to be dominated by 
their known $T=0$ forms up to the inter-monopole separation, whereupon they 
match onto the asymptotics of the paramagnetic 
phase~\cite{footnote:dipolar}. 
Secondly, the low temperature thermodynamics of spin ice can be transformed 
from that of a dense set of \emph{localised} dipolar spins to that of a 
dilute set of \emph{itinerant} Coulombically interacting particles---a 
(magnetic) Coulomb liquid as first noted in Ref.~\onlinecite{Castelnovo2008}: 
\beq
H
=
\frac{\mu_0}{4 \pi} \sum_{i < j}
  \frac{q_i q_j}{r_{ij}}
+ 
\Delta \sum_{i}
  \left( \frac{q_i}{2\mu/a_d} \right)^2
,
\nonumber 
\eeq
where the charges $q_i$ take the values $\pm 2 \mu / a_d$, 
$\mu \simeq 10 \mu_B$ being the dipole moment of a spin and $a_d$ the distance 
between the centres of adjacent tetrahedra (diamond lattice constant in 
Fig.~\ref{fig: lattice details}), and $\Delta$ is the energy cost of a 
monopole. 

The transformation is extremely helpful as much is known about 
Coulomb liquids, with a venerable history spanning fields 
from statistical physics all the way to the chemistry of electrolytes.
Indeed, the known properties of the Coulomb liquid have led to an explanation
of the `liquid-solid' phase transition of spin ice in a $[111]$ 
field~\cite{Castelnovo2008}, as well as of its magnetic specific 
heat~\cite{Morris2009} in zero field. 
More recently, much attention has been devoted to the study of the
``magnetricity''~\cite{Bramwell2009} in these 
``magnetolytes''~\cite{CastelnovoCPC}, the equilibrium and non-equilibrium 
behaviour of such a magnetic Coulomb liquid, inspired by the analogous 
electric phenomena such as the Wien effect~\cite{Bramwell2009,Giblin2010}. 

In this paper, expanding on our previous work in Ref.~\onlinecite{Morris2009}, 
we develop a low-energy theory for spin ice in the framework of
the Debye-H\"uckel (DH) theory of a dilute Coulomb liquid. DH theory will be 
familiar to readers from many different disciplines but to our knowledge has 
never been applied to a three-dimensional magnetic material before the advent 
of spin ice.

The purpose of this paper is two-fold. First, it gives a detailed and
context-specific account of the DH theory for spin ice. Second, its ability
to model experimental data is underlined. In particular, we show that an
existing framework to describe the dynamics of spin ice, when supplemented by
DH theory, provides improved agreement with existing experimental and
numerical data on the AC-susceptibility of spin
ice~\cite{Matsuhira2000,Matsuhira2001,Snyder2004,Ryzhkin2005,Jaubert2009}. 

This is perhaps as good a point as any to digress and address the concerns of 
readers who may be worried that our replacement of spins by monopoles is too 
good to be true. 
Here three points are in order. First, as we have already noted above, the 
spins do enter the static correlations but in a manner that is understood. 
Second, a given monopole configuration can be ``dressed'' by many spin 
configurations. However summing over these dressings generates an 
effective entropic Coulomb attraction between the monopoles at long 
wavelengths (see e.g., Ref.~\onlinecite{Henley_coulomb}) which can {\it also} 
be included in the Coulomb/DH framework. We will address this point is 
Sec.~\ref{sec: spin entropy effects} and find that the entropic effect can be 
ignored for the present set of spin ice compounds. Third, there is still a 
remaining issue that not all monopole configurations are in fact compatible 
with some spin configuration, and moreover the spins can induce non-trivial 
structure to the monopole energy landscape which in turn can significantly 
alter dynamical properties of spin ice out of 
equilibrium~\cite{Castelnovo2010}. 
However, these are weak constraints on the Coulomb framework 
and it seems highly unlikely that they play any role in determining 
equilibrium properties.

\begin{figure}[ht]
\vspace{0.2 cm}
\begin{center}
\includegraphics[width=0.99\columnwidth]
                {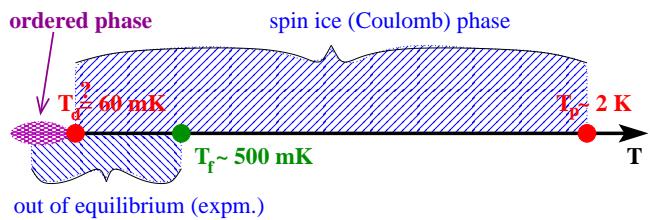}
\end{center}
\caption{
\label{fig: SI temperatures}
Schematic illustration of the different temperature regimes in spin ice,
separated by $T_d$, $T_f$, and $T_p$ as explained in the text.
The putative ordering below $T_d$ appears to be prevented by freezing of the 
magnetic degrees of freedom below $T_f$, as evidenced e.g., by a discrepancy 
between field-cooled and zero-field-cooled magnetisation. 
At temperatures of about $T_p$, the materials cross over to a trivial 
paramagnetic behaviour.
}
\end{figure}

We close the introduction by remarking on the range of applicability of the 
Coulomb liquid/DH theory framework in the actual compounds 
(see Fig.~\ref{fig: SI temperatures}). 
At high temperatures, above a scale $T_p$, we are in a conventional 
paramagnetic regime where the monopoles are dense. 
Below $T_p$ the monopoles become sufficiently dilute that they can be treated 
by DH theory. At a much lower temperature $T_d$, the Coulomb phase is 
unstable to ordering transitions~\cite{SIordering,Melko2001,Yavorskii2008}, 
the details of which are not entirely settled. For the model Hamiltonian, 
$T_d \equiv 0$.
While the Coulomb liquid framework should thus apply in the range 
$T_d < T < T_p$, the equilibrium DH treatment runs into problems around a 
temperature $T_f > T_d$ where the system falls out of equilibrium before 
any ordering is visible.
Much of the interest in the spin ice compounds {\DTO} and {\HTO} derives from
the fact that $T_d,T_f<T_p$, so that there is a window where Coulomb physics
is well visible.
%
%

The remainder of this paper is organised as follows: we first provide DH
background, discuss specificities of its application in the spin ice setting,
discuss its range of validity and finally apply it to experiment. In addition,
we discuss two other topics of import in this context. Firstly, we determine
the size of the entropic Coulomb interaction between monopoles. Secondly, we
compute the low-temperature mobility of magnetic monopoles in spin ice with a
single-spin flip dynamics believed to be appropriate for experimental
compounds {\DTO} and {\HTO}.
%
%

\section{\label{sec: DH free energy}
Debye-H\"{u}ckel free energy
        }
We now turn to the application of DH theory to spin ice.
The reader not interested in details of the formalism can skip ahead
to Section~\ref{sec: simulations}.
%
%

\subsection{\label{sec: non-interacting limit}
Non-interacting monopoles
           }
To lay the foundation, let us start by considering the simple case of
non-interacting monopoles, corresponding to a nearest-neighbour spin ice model.
Since the monopole description of spin ice is valid only when the density of
defective tetrahedra is sufficiently small, i.e., at low temperatures,
we consider only the less costly defects (3in-1out and 3out-1in
tetrahedra) and neglect charge 2 excitations altogether (4in-0out and
4out-0in) as they cost four times as much energy. The internal energy $U$ of
the system is thus proportional to the number of monopoles $N$,
\begin{eqnarray}
U = N \Delta = N_t \rho \Delta
\ ,
\end{eqnarray}
where $\Delta$ is the energy cost of an isolated monopole (assumed in the
following to be measured in Kelvin) and $\rho \equiv N/N_t$ is
the monopole density per tetrahedron.

The number of configurations that an ensemble of $N/2$ positive (hard-core)
monopoles and $N/2$ negative ones can take on a lattice of $N_t$ sites
($N_t$ being the total number of tetrahedra in the system) is given by
\begin{eqnarray}
W = \left({N_t\atop{N/2 \ N/2 \ (N_t-N)}}\right)
.
\end{eqnarray}
Using Stirling's approximation in the large $N_t$ and large $N$ limit,
we obtain the $\mathcal{S} = k_B \ln W$ `entropy of mixing',
\begin{eqnarray}
S &\equiv& \mathcal{S}/k_B 
\nonumber \\ 
&=& 
-N_t \left[2 (\rho/2) \ln \left(\rho/2\right) + (1-\rho) \ln(1-\rho) \right]
\label{eq: mixing entropy}
\end{eqnarray}
with a concomitant free energy per spin
\begin{equation}
\frac{F_{\rm nn}}{N_s k_B}  =
\frac{U - T S}{N_s}
\label{eq: free energy nn}
\end{equation}
where the number of spins is twice the
number of tetrahedra, $N_s = 2 N_t$.
Minimizing with respect to $\rho$,
we obtain the known expression for the total monopole density
\bea
\rho_{\rm nn}
&=&
\frac{2\exp(-\Delta/T)}{1+2\exp(-\Delta/T)}
.
\label{eq: rho nn}
\end{eqnarray}
For small $T$, and hence small $\rho_{\rm nn}$,
$\rho_{\rm nn} \simeq 2 \exp(-\Delta/T)$.
For large $T$, Eq.~\eqref{eq: rho nn} tends asymptotically to the value
$2/3$, which is clearly incorrect -- as expected since random Ising spins
on a pyrochlore lattice yield a density $\rho_{\textrm{random}} = 5/8$ of
defective tetrahedra. This can be seen e.g., if we consider a single 
tetrahedron: out of the $2^4 = 16$ allowed Ising configurations, only $6$ 
satisfy the 2in-2out condition and the remaining $10$ configurations violate 
charge neutrality. 

%
%

\subsection{\label{sec: DH contribution}
Debye-H\"{u}ckel contribution
           }
One of the major approximations in Sec.~\ref{sec: non-interacting limit}
is the fact that the long range Coulomb interactions
between the monopoles were entirely neglected~\cite{Castelnovo2008}. 
Taking advantage of the analogy between spin ice defects and a two-component
Coulomb liquid (in the absence of appplied magnetic fields), we can use the
Debye approximation to estimate the magnetostatic contribution
to the free energy (in degrees Kelvin per spin):~\cite{Debye1923}
\bea
\frac{F_{\rm el}}{N_s k_B}
&=&
-
\frac{N T}{4 N_s \pi \rho_V a^3_d}
\left[
  \frac{(a_d \kappa)^2}{2} - (a_d \kappa) + \ln(1 + a_d \kappa)
\right]
\nonumber \\
\kappa
&=&
\sqrt{\frac{\mu_0 q^2 \rho_V}{k_B T}}
,
\eea
where $\rho_V = N / V$ is the dimensionful volume density of monopoles and
$a_d$ is the distance between the centres of two neighbouring tetrahedra
(i.e., the dual diamond lattice constant).

It is convenient to express the dimensionless quantity $a_d \kappa$ in terms
of the Coulomb energy between two neighbouring monopoles
$E_{\rm nn} \equiv \mu_0 q^2 / (4 \pi a_d \, k_B)$,
\bea
a_d \kappa
&=&
\sqrt{4 \pi}
\sqrt{\frac{E_{\rm nn}}{T} (\rho_V a^3_d)}
.
\eea
Here $q$ stands for the magnitude of the monopole charge ($q = 2\mu/a_d$,
where $\mu$ is the rare earth magnetic moment~\cite{Castelnovo2008}).

There are $8$ diamond lattice sites in a $16$-spin cubic unit
cell of side $(4/\sqrt{3}) \, a_d$. The total volume of the system can then
be written as $V = (N_t/8) (4/\sqrt{3})^3 \, a^3_d$ and
\beq
\rho_V a^3_d
=
\frac{N}{V/a^3_d}
=
\frac{3\sqrt{3}}{8} \rho
.
\eeq
As a result, we arrive at
\bea
\frac{F_{\rm el}}{N_s k_B}
&=&
-
\frac{T}{3 \sqrt{3} \pi}
\left[
  \frac{(a_d \kappa)^2}{2} - (a_d \kappa) + \ln(1 + a_d \kappa)
\right]
\nonumber \\
\label{eq: magnetostatic energy}
\\
a_d \kappa
&=&
\sqrt{\frac{3 \sqrt{3} \pi E_{\rm nn}}{2 T}}
\,
\sqrt{\rho}
\equiv
\alpha(T) \sqrt{\rho}
,
\label{eq: kappa debye}
\eea
where the last equation defines the function $\alpha(T)$.
In the low temperature limit, the magnetostatic contribution
scales as $\rho^{3/2}$, namely
\bea
\frac{F_{\rm el}}{N_s k_B}
&\simeq&
-
\frac{T}{3 \sqrt{3} \pi}
  \frac{(a_d \kappa)^3}{3}
\nonumber \\
&\simeq&
-
\sqrt{\frac{\pi}{8 \sqrt{3}}}
  E_{\rm nn} \sqrt{\frac{E_{\rm nn}}{T}} \rho^{3/2}
.
\eea

We can then combine Eqs.~\eqref{eq: magnetostatic energy}
and~\eqref{eq: kappa debye} with Eq.~\eqref{eq: free energy nn} from
Sec.~\ref{sec: non-interacting limit} to obtain a mean field free energy
-- per spin in degrees Kelvin -- of an ensemble of $N$ monopoles on a
lattice with long range Coulomb interactions:
\begin{eqnarray}
\frac{F}{N_s k_B}
&=&
\frac{\rho}{2} \Delta
+
\frac{T \rho}{2} \ln\left(\frac{\rho/2}{1-\rho}\right)
+
\frac{T}{2} \ln(1-\rho)
\nonumber \\
&-&
\frac{T}{3 \sqrt{3} \pi}
\left\{
  \frac{\alpha^2(T) \, \rho}{2}
  - \alpha(T) \sqrt{\rho}
  + \ln\left[1 + \alpha(T) \sqrt{\rho}\right]
\right\}
\label{eq: free energy DH}
\nonumber \\
\alpha(T)
&=&
\sqrt{\frac{3 \sqrt{3} \pi E_{\rm nn}}{2T}}
. 
\label{eq: free en free monopoles}
\end{eqnarray}
Note that this reduces to the non-interacting limit if we set
$E_{\rm nn} = 0$.

Minimizing with respect to the defect density $\rho$, one obtains a
self-consistent set of equations:
\bea
\frac{d (F/N_s k_B)}{d \rho}
&=&
\Delta + T \ln\left(\frac{\rho/2}{1-\rho}\right)
-
\frac{E_{\rm nn}}{2}
\frac{\alpha(T)\sqrt{\rho}}
     {1+\alpha(T)\sqrt{\rho}}
=
0
\nonumber \\
\rho
&=&
\frac{2\exp\left[-\left(
                    \frac{\Delta}{T}
		    -
		    \frac{E_{\rm nn}}{2T} \frac{\alpha\sqrt{\rho}}{1+\alpha\sqrt{\rho}}
		  \right)\right]}
     {1+2\exp\left[-\left(
                    \frac{\Delta}{T}
		    -
		    \frac{E_{\rm nn}}{2T} \frac{\alpha\sqrt{\rho}}{1+\alpha\sqrt{\rho}}
		  \right)\right]}
.
\label{eq: self-consistent rho}
\eea

Unfortunately, Eq.~\eqref{eq: self-consistent rho} cannot be solved
analytically and one has to resort to numerical methods to obtain $\rho(T)$.
We find that the recursive approach
\bea
\rho_0 &=& \rho_{\rm nn} = \frac{2\exp(-\Delta/T)}{1+2\exp(-\Delta/T)}
\nonumber \\
\rho_{\ell+1} &=&
\frac{2\exp\left[-\left(
                    \frac{\Delta}{T}
		    -
		    \frac{E_{\rm nn}}{2T}
		    \frac{\alpha\sqrt{\rho_{\ell}}}{1+\alpha\sqrt{\rho_\ell}}
		  \right)\right]}
     {1+2\exp\left[-\left(
                    \frac{\Delta}{T}
		    -
		    \frac{E_{\rm nn}}{2T}
		    \frac{\alpha\sqrt{\rho_{\ell}}}{1+\alpha\sqrt{\rho_\ell}}
		  \right)\right]}
\label{eq: DH recursive solution}
\eea
converges with acceptable accuracy in less than $5$ iterations.
Substituting $\rho \equiv \rho_{\ell \to \infty} \simeq \rho_{5}$ into
Eq.~\eqref{eq: free energy DH} we obtain numerically the approximate free
energy of dipolar spin ice as a function of temperature.

Between Eqns. (\ref{eq: free energy DH}) and (\ref{eq: self-consistent rho}) 
we have obtained the free energy for monopoles in the DH approximation. 
From this one can compute several thermodynamic quantities of interest 
(see e.g., Sec.~\ref{sec: DH secific heat}).
%
%

\section{\label{sec:paracons}
Spin ice parameters and DH internal consistency
        }
We first derive the parameters describing the {\DTO} and {\HTO} spin ices
within the dumbbell model~\cite{Castelnovo2008} in the subsequent subsection.
Following the determination of the parameters, we discuss the range of
temperatures over which the treatment is valid.
%
%

\subsection{\label{sec: parameters}
Spin ice parameters in the dumbbell model
           }
The usefulness of the dumbbell model lies in the fact that it correctly
captures the long-distance form of the dipolar interaction -- as well as the
magnetic Coulomb  interaction between the monopoles -- while preserving the
degeneracy of the spin ice states. At the same time, a model of such
simplicity cannot do justice to the full short-distance structure of the
interactions present in the real compound, which include further-neighbour
superexchange as well as quadrupolar interaction terms between the spins.
We will thus find in the following sections that the best fit to both
numerics and experiment requires slight adjustments to the dumbbell model
parameters to obtain {quantitatively} optimal fits.

We also take this opportunity to caution the reader that the 'microscopic'
parameters themselves are subject to change on the level of a few percent
as experiments and their detailed numerical modeling evolve 
(and, hopefully, improve) over time. Such changes can be innocuous
(e.g. a 1\% change to the diamond lattice constant) but since some of the
resulting physics is rather delicate, they can feed through to relatively
larger corrections, most prominently as a factor $3$ in the estimated value
of $T_d$!~\cite{Melko2001,Yavorskii2008}

From the pyrochlore lattice constant $a = 3.54$~{\AA} one obtains the diamond
lattice constant $a_d = \sqrt{3/2}\,a = 4.34$~\AA.
Combined with the spin magnetic moment $\mu = 10 \mu_B$
($\mu_B = 9.27\,10^{-24}$ J/Tesla), this gives the monopole charge
$q \simeq 4.6\:\mu_B / \textrm{\AA} \simeq 4.28\,10^{-13}$~J/(Tesla\,m)
(see Ref.~\onlinecite{Castelnovo2008} and Supplementary Information therein). 

Inserting the dipolar coupling constant
\bea
D = \frac{\mu_0}{4\pi k_B} \frac{\mu^2}{a^3} \simeq 1.41 \,\textrm{K}
\nonumber
\eea
($\mu_0 / 4\pi = 10^{-7}$~N/A$^2$, $k_B = 1.38\,10^{-23}$~J/K)
and the nearest-neighbour exchange coupling $J \simeq -3.72$~K
for Dy$_2$Ti$_2$O$_7$ ($J \simeq -1.56$~K for Ho$_2$Ti$_2$O$_7$) into the
expression for the bare cost of a single isolated monopole in
Ref.~\onlinecite{Castelnovo2008}, we obtain
\bea
\Delta
&=&
\frac{1}{2} v_0 q^2
=
\frac{2J}{3} + \frac{8}{3}\left[ 1 + \sqrt{\frac{2}{3}} \right]\,D
\label{eq: Delta}
\\
&=&
\begin{cases}
4.35\,\textrm{K}
&
\textrm{for Dy$_2$Ti$_2$O$_7$} \; (J=-3.72\,\textrm{K})
\\
5.79\,\textrm{K}
&
\textrm{for Ho$_2$Ti$_2$O$_7$} \; (J=-1.56\,\textrm{K})
\end{cases}
.
\nonumber
\eea
%
%
The energy of two monopoles at nearest neighbour distance is:
\bea
E_{\rm nn}
&=&
\frac{\mu_0}{4\pi k_B}\frac{q^2}{a_d}
\simeq
3.06\,\textrm{K}
.
\eea
Therefore, the creation of two neighbouring monopoles by a single spin
flip event in a spin ice configuration where all tetrahedra satisfy the
2in-2out rules incurs an energy cost
\beq
\Delta_s
=
2 \Delta - E_{\rm nn}
\simeq
\begin{cases}
5.64 \,\textrm{K} & \textrm{for Dy$_2$Ti$_2$O$_7$}
\\
8.52\,\textrm{K} & \textrm{for Ho$_2$Ti$_2$O$_7$}
\end{cases}
.
\eeq

As a final remark, it is interesting to compare the
force between two monopoles at nearest neighbour distance,
\bea
F_{\rm nn}
&=&
\frac{\mu_0}{4\pi}\frac{q^2}{a^2_d}
\simeq
9.74\,10^{-14}\,\textrm{N}
,
\eea
to that between two eletrons at the same distance,
$F_\textrm{el} \simeq 1.22\,10^{-9}$\,\textrm{N}, four orders of magnitude
stronger! By contrast, a pair of Dirac monopoles would experience a force of
almost $10^{-5}$N.
%
%

\subsection{\label{sec: DH screening length}
Internal consistency: screening length vs. monopole separation and lattice
constant
           }
The Debye screening length $\xi_{\rm Debye}$ is given by the inverse
of the constant $\kappa$ in Eq.~(\ref{eq: kappa debye}). In units of the
diamond lattice constant $a_d$ this amounts to
\beq
\frac{\xi_{\rm Debye}}{a_d}
=
\frac{1}{a_d \kappa}
=
\sqrt{\frac{2 T}{3 \sqrt{3} \pi E_{\rm nn}}}
\,
\frac{1}{\sqrt{\rho}}
.
\label{eq: DH screening length}
\eeq

\begin{figure}[ht]
\vspace{0.2 cm}
\begin{center}
\includegraphics[width=0.99\columnwidth]
                {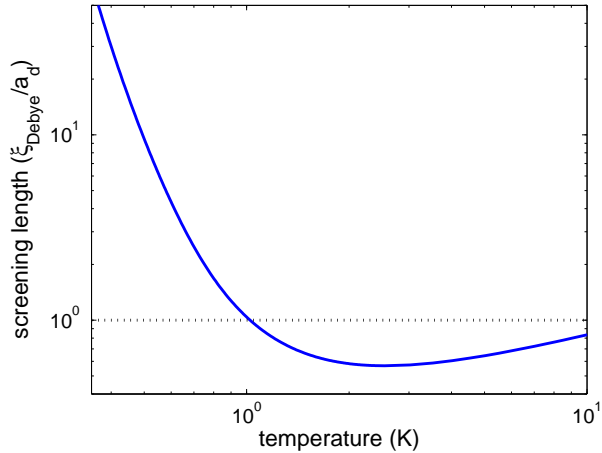}
\end{center}
\caption{
\label{fig: debye screening}
Plot of the Debye screening length vs temperature, using the density from
the numerical solution to the Debye-H\"{u}ckel calculation in
Sec.~\ref{sec: DH contribution}.
}
\end{figure}

The dependence of $\xi_{\rm Debye} / a_d$ on temperature, after substituting
$\rho(T)$ from the numerical solution of Eq.~\eqref{eq: self-consistent rho}
is illustrated in Fig.~\ref{fig: debye screening}
(using for instance $\Delta = 4.7$~K).

We anticipate here that there is a systematic discrepancy between the DH 
approximation and the MC simulation results on the heat capacity for 
$T \gtrsim 1$~K (see Fig.~\ref{fig: MC spec heat}). 
To understand this, we note the following.

Firstly, above $T \simeq 1$~K the screening length becomes shorter than
the lattice spacing. This artefact arises because the DH term
in the free energy was derived in the continuum.
For $T \gtrsim 1$~K one thus needs to consider the DH results with caution.
Having said this, once the screening length gets very short, the long range
nature of the Coulomb interaction becomes less important. One can then
reliably truncate the interactions to short range and use alternative
approaches to compute the free energy and other thermodynamic quantities, as
illustrated for instance in Appendix~\ref{app: single tet approx}.

Secondly, as $T$ approaches the Curie-Weiss temperature of about 2K, the
average separation between monopoles, $d \sim a_d\,\rho^{-1/3}$, becomes
comparable to the lattice constant $a_d$ and the monopole picture is no longer
appropriate to describe spin ice -- monopoles are useful as long as they are
sparse, otherwise it is more efficient to work directly with the microscopic
spin degrees of freedom.
(In addition, for even higher values of $T$, the neglect of doubly-charged
monopoles becomes problematic.)
For instance, it would be more appropriate to use a conventional
high-temperature series expansion.

Another parameter of physical relevance is the ratio of screening length to
monopole separation: the larger this ratio, the more appropriate a continuum
description is.
The dimensionful monopole density $\rho_V$ can be expressed in terms
of the monopole density per tetrahedron $\rho$ (which appears in the DH
calculations in Sec.~\ref{sec: DH free energy}) using the relation
$\rho_V = 3 \sqrt{3} \rho / (8 a_d^3)$.
From it, we can obtain the average monopole separation $\rho_V^{-1/3}$.
By comparing these two length scales, one observes that DH theory is near
an `internal' limit of validity, as the ratio
$\xi_{\rm Debye} / \rho_V^{-1/3}$ is close to one throughout
the range of interest.
Indeed, $\xi_{\rm Debye} / \rho_V^{-1/3} \gtrsim 1$ only below $300$~mK,
dropping by a factor three towards its minimum at $1$~K (not shown).
%
%

\subsection{\label{sec: DH vs noint}
Role of the magnetostatic contribution
           }
\begin{figure}[ht]
\vspace{0.2 cm}
\begin{center}
\includegraphics[width=0.99\columnwidth]
                {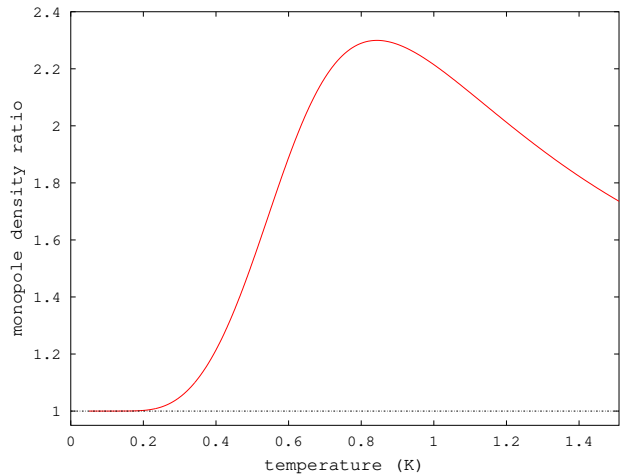}
\end{center}
\caption{
\label{fig: DH vs noint}
Ratio of the monopole densities from Sec.~\ref{sec: DH contribution} 
obtained with and without the Debye-H\"{u}ckel magnetostatic contribution, 
Eq.~\eqref{eq: magnetostatic energy}, as a function of temperature.
}
\end{figure}

It is interesting to quantify how big the change brought about by the DH 
accounting of Coulomb interactions and screening actually is. 
To do this, let us consider the density of monopoles, which will play a role 
later in the comparison with Monte Carlo simulation results 
(Sec.~\ref{sec: DH mono density}). 
In Fig.~\ref{fig: DH vs noint} we plot the ratio of the monopole densities 
from Sec.~\ref{sec: DH contribution} with and without the magnetostatic 
contribution Eq.~\eqref{eq: magnetostatic energy}, 
using parameters appropriate for spin ice {\DTO}. 
%
%
%
Within the region $T \lesssim 1$~K, one notices that DH theory can lead to 
a more than two-times larger monopole density. 
Given that spin ice materials are prone to falling out of 
equilbrium at temperatures $T \lesssim 0.5$~K, the behaviour of the system 
in the temperature window where DH corrections are sizeable is of crucial 
relevance to experiment. 
In the limit of low temperatures, the DH correction instead becomes less and 
less important. 
%

\subsection{\label{sec: pairing}
Monopole-antimonopole pairing } Debye-H\"{u}ckel theory neglects the
association of monopoles into neutral dipolar pairs (see
Ref.~\onlinecite{Fisher1993} and references therein).  Although this
can in general lead to sizeable discrepancies between DH predictions
and experiments, we argue hereafter that pairing corrections are small
for the observables in spin ice that we consider here, due to the
combination of its limit of validity ($T \lesssim 1$~K, see
Sec.~\ref{sec: DH screening length}) and the relatively 
larger energy cost for a
monopole excitation, $\Delta \sim 4-5$~K, in comparison to the Coulomb
energy when hard core charges come into ``contact'' (nearest-neighbour
distance), $E_{\rm nn} \simeq 3.06$~K.

In order to show this, let us assume that monopoles in spin ice are either 
free (density $\rho_0$), if separated by a distance larger than $\ell_B$, 
or bound in a pair, if separated by a distance $d$ shorter than $\ell_B$. 
Here we choose $\ell_B$ to equal the Bjerrum length, at which the thermal
energy $k_B T$ equals the Coulomb energy:
\beq
\ell_B/a_d 
= 
\frac{\mu_0}{4\pi a_d} \frac{\left( 2\mu/a_d \right)^2}{2 k_B T} 
\simeq 
\frac{1.54}{T\mathrm{[K]}} \quad \textrm{for {\DTO}}
, 
\eeq

We now consider only Coulomb interactions amongst free monopoles 
and between the two monopoles belonging to the same pair, while we neglect 
monopole-pair and pair-pair interactions, on the grounds that they are 
generally weaker and they decay faster with distance. 
We also neglect excluded volume effects (therefore, any 
results we obtain ought to be treated with care as the density of monopoles 
approaches unity, which is anyway not the regime we are interested in). 

The free energy $f_0$ for the fraction of free monopoles in the system is 
straightforwardly given by Eq.~\ref{eq: free en free monopoles}. 
The potential energy term for the bound pairs, of densities $\rho_d$, 
$d=1,2,\ldots,\ell_B$, is also immediate to write as it involves only the 
inter-pair Coulomb term: $(2\Delta - E_d)\rho_d$, where 
$E_d \sim E_{\rm nn}/d$. 
The entropic contribution to the free energy of a bound pair of 
characteristic distance $d$ can be computed from the numbers of ways that 
such pair can appear on the lattice, 
\bea
W &=& \left( {N_t}\atop{N_t \rho_d} \right) v_d^{N_t \rho_d}
\\ 
\frac{S}{N_t k_B} &=& \ln W 
\nonumber \\
&=& 
- \rho_1 \ln \left(\rho_1\right) - (1-\rho_1) \ln(1-\rho_1) 
\nonumber \\
&& + \rho_1 \ln(v_d)
, 
\eea
where $v_d$ is the number of configurations that the two monopoles in the 
pair can take, given say that the centre of mass of the pair is fixed. 
For a nearest-neighbour pair, $v_1 = 2$. 
For large values of $d$, we expect $v_d$ to scale as $2 \times 4 \pi d^2$. 
In practice, we shall approximate 
\beq
v_d = v_1 \frac{8 \pi d^2}{8 \pi (d=1)^2} 
= v_1 d^2
= 2 d^2 
. 
\eeq

Combining these results, we obtain the free energies (per tetrahedron) for 
free and bound pairs, 
\bea
f_0 &=& 
\frac{F_{\rm el}}{N_t k_B} + \Delta \rho_0 
\nonumber\\
&+& 
T \left[ \rho_0 \ln \left(\rho_0/2\right) + (1-\rho_0) \ln(1-\rho_0) \right]
\\
f_d 
&=& 
(2\Delta - E_d) \rho_d 
\nonumber\\
&+& 
T \left[ \rho_d \ln \left(\rho_d\right) + (1-\rho_d) \ln(1-\rho_d) \right]
\nonumber\\
&-& 
T \rho_d \ln(v_d)
, 
\label{eq: free energies pairs}
\eea
as a function of the densities $\rho_0$ and $\rho_d$, $d=1,\ldots,\ell_B$. 
The equilibrium free energy of the entire system is then obtained minimizing 
the sum 
\begin{eqnarray}
f_{\rm tot} = f_0 + f_1 + \ldots + f_{\ell_B} 
\nonumber 
\eea
with respect to $\rho_0$, $\rho_1$, \ldots, $\rho_{\ell_B}$. 

Unlike  $\rho_0$,  already considered in 
Sec.~\ref{sec: DH contribution}, the $\rho_d$ are obtained
straigthforwardly as
\bea
\rho_d &=& 
\frac{v_d \, e^{-(2\Delta - E_d)/T}}{1 + v_d \, e^{-(2\Delta - E_d)/T}}
. 
\label{eq: rho_d}
\eea
Clearly, an intrinsic limit of validity of the theory is given by the 
condition that 
\beq
\rho_{\rm tot} \equiv \rho_0 + 2 \sum_{d=1}^{\ell_B} \rho_d \leq 1 
. 
\eeq

In addition, we are of course in particular interested in 
$\rho_0\gg 2 \sum_{d=1}^{\ell_B}\rho_d=\rho_b$. 

The behaviour of $\rho_0$, $\rho_1$, $\rho_{b}$ and $\rho_{\rm tot}$ 
as a function of temperature in the regime of interest to spin ice is 
shown in Fig.~\ref{fig: bound pair density}. 
\begin{figure}[ht]
\vspace{0.2 cm}
\begin{center}
\includegraphics[width=0.99\columnwidth]
                {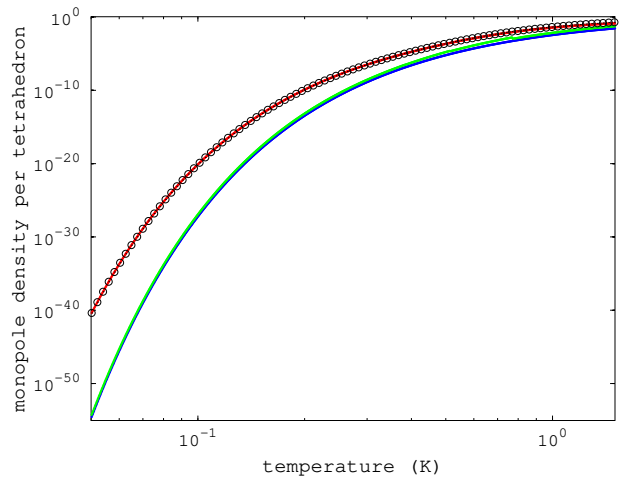}
\end{center}
\caption{
\label{fig: bound pair density} 
Behaviour of $\rho_0$ (red), $\rho_1$ (blue), $\rho_{b}$ (green), and 
$\rho_{\rm tot}$ (open black circles). In the regime of interest to spin 
ice physics, the total monopole density is dominated, 
\emph{at equilibrium}, by the free monopoles. 
}
\end{figure}
While at $T=1$~K the bound pairs make up for approximately $16$\% of the 
monopoles in the system, this quickly drops to $7$\% at $T=500$~mK and  to 
$\lesssim 10^{-5}$\% for $T \lesssim 100$~mK. 

Of course, all the considerations in this section apply when the 
system is in thermal equilibrium. This is known not to be always the case 
in experimental settings involving spin ice materials! For example, 
as discussed in Ref.~\onlinecite{Castelnovo2010}, fast variations in the 
temperature of a sample can lead to a ``population inversion'', whereby a 
relatively high density of monopoles survives \emph{out of equilibrium} 
down to very low temperatures, mostly forming nearest-neighbouring pairs 
($\rho_{\rm tot} \simeq \rho_1$)~\cite{Castelnovo2010}. 

The arguments presented in this section are akin to the so-called Bjerrum 
correction to DH. The latter typically leads, at low temperatures, to the 
condensation of all monopoles into bound pairs. This is an artifact due to 
the neglecting of monopole-pair interactions, as discussed in 
Ref.~\onlinecite{Fisher1993}. 

Our results do not exhibit any such condensation.  The reason for this
difference in behaviour are to be found in the large monopole cost
with respect to the Coulomb energy at nearest-neighbour distance.  The
net energy gain in the formation a bound pair is insufficient to
compensate for the corresponding entropy loss.  The situation would be
dramatically different if the creation cost of the monopoles were
lowered such that it can be offset by the Coulomb attraction to
another monopole.

For completeness, we mention that for sufficiently large Coulomb
attraction the chemical potential of a bound pair would have the
\emph{opposite sign} with respect to that of a free monopole, leading
to a collapse of the system into an ionic crystal of monopoles. In
spin language, this tranlates into an instability of spin ice to an
ordered ground state.
%
%

\section{\label{sec: simulations}
Comparison of DH with Monte Carlo
        }
We compare the DH results above with Monte Carlo (MC) simulations
using the spin ice parameters in Ref.~\onlinecite{Melko2004},
reported in the previous section.
The Ewald summation technique was used for the long range dipolar
interactions between the spins~\cite{Ewald_refs}.
We used systems of size $16 L^3=3456$ spins ($L=6$)
and single spin flip updates.
%
%

\subsection{\label{sec: DH mono density}
Monopole density
           }
A first comparison between the non-interacting limit and the DH approach can
be done by looking at the resulting monopole density as a function of
temperature, Eq.~\eqref{eq: rho nn} and the numerical solution
to~\eqref{eq: self-consistent rho}, illustrated in
Fig.~\ref{fig: monopole density} together with the monopole density from
Monte Carlo simulations of dipolar spin ice.
\begin{figure}[ht]
\vspace{0.2 cm}
\begin{center}
\includegraphics[width=0.99\columnwidth]
                {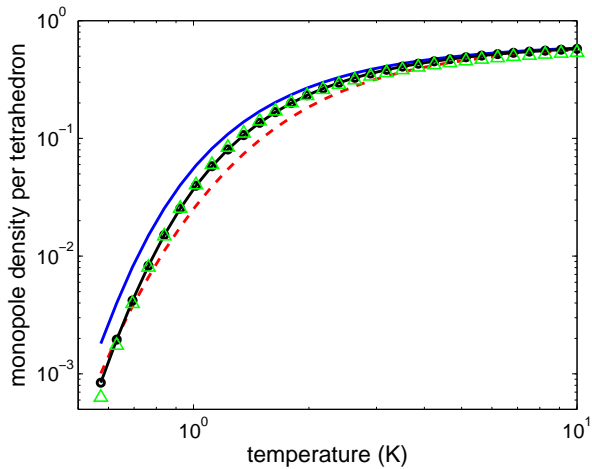}
\end{center}
\caption{
\label{fig: monopole density}
Monopole density from numerical simulations (green triangles), compared to the
analytical result in the non-interacting approximation (dashed red line) and
in the DH approximation (solid blue line).
Note that there are no fitting parameters.
An improved agreement between the simulations and the DH
approximation obtains if we adjust the bare monopole cost to
$\DeltaMC = 4.7$~K (black dotted curve).
}
\end{figure}

The agreement between DH and MC results  is already quite reasonable  yet it
improves considerably if we tune the bare monopole cost to
$\DeltaMC = 4.7$~K.  As mentioned above, we believe the origin of this
adjustment to be in the short-distance physics beyond the dumbbell model of
Ref.~\onlinecite{Castelnovo2008}.
In quantities sensitive to such short range details, such as
$\Delta$, this 8\% discrepancy is not unreasonable.
%
%

\subsection{\label{sec: DH vs MC heat capacity}
Heat capacity
           }
Given the DH free energy (expressed in units of degree Kelvin per Dy ion),
one can obtain the heat capacity of the system in units of
J~mol$^{-1}$K$^{-1}$ via the thermodynamic relation
\bea
c_V
&=&
- N_A k_B T \, \partial^2_T (F/N_s k_B)
,
\label{eq: spec heat from free energy}
\eea
where
$N_A$ is Avogadro's number,
$\beta=1/k_B T$, and $k_B$ is the Boltzmann constant.

In MC simulations, $c_V$ can be obtained by the usual fluctuation-dissipation
route, measuring the average energy $\langle \varepsilon \rangle$ and
its fluctuations,
\bea
c_V
&=&
\frac{R N_s}{T^2}
\left[
  \langle \varepsilon^2 \rangle - \langle \varepsilon \rangle^2
\right]
.
\label{eq: spec heat from energy fluct}
\eea

A comparison between the non-interacting calculations, Eq.~\eqref{eq: rho nn}
and Eq.~\eqref{eq: free energy nn}, the DH calculations,
Eq.~\eqref{eq: self-consistent rho} and Eq.~\eqref{eq: free energy DH},
the single tetrahedron approximation in Appendix~\ref{app: single tet approx},
and Monte Carlo simulations is shown in Fig.~\ref{fig: MC spec heat}.
\begin{figure}[ht]
\vspace{0.2 cm}
\begin{center}
\includegraphics[width=0.99\columnwidth]
                {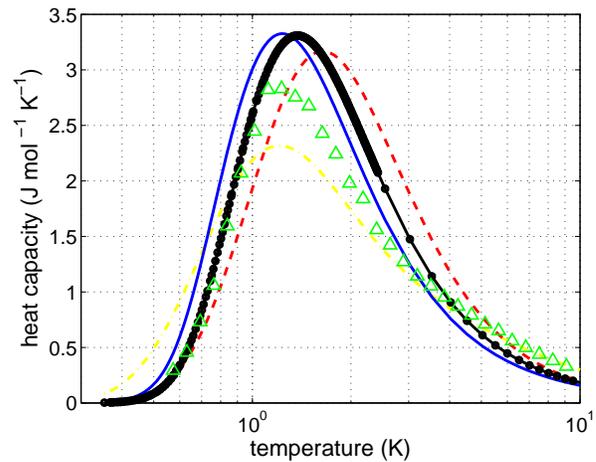}
\end{center}
\caption{
\label{fig: MC spec heat}
Heat capacity from numerical simulations (green triangles), compared to the
analytical result in the non-interacting approximation (dashed red line) and
in the DH approximation (solid blue line).
Note that there are no fitting parameters.
Like for the density (cf. Fig.~\ref{fig: monopole density}), improved
agreement between the simulations and the DH solution is obtained for a
bare monopole cost $\DeltaMC = 4.7$~K (black dotted curve).
The single-tetrahedron approximation discussed in
Appendix~\ref{app: single tet approx} can only be made to agree with the
experimental results on a very narrow temperature range, even if we use
$J_{\rm eff}$ as a fitting parameter (dash-dotted yellow line).
}
\end{figure}

Consistently with the monopole density results, a comparison of the
heat capacity from DH theory and simulations also shows improved
agreement using $\DeltaMC=4.7$~K instead of $\Delta = 4.35$~K.
We shall see in Sec.~\ref{sec: Comparison to experiment} that an 8\% larger 
value of $\Delta$ with respect to Eq.~\eqref{eq: Delta} is also consistent 
with the comparison between DH theory and experimental results. 

The results in Fig.~\ref{fig: monopole density} and
in Fig.~\ref{fig: MC spec heat} clearly show that:
(i) a theory of point-like Coulomb-interacting charges
(in particular with the improved value of the bare monopole cost)
goes a long way into
capturing the physics of spin ice, much better than conventional approaches
based on truncated cluster expansions of the free energy of the system;
(ii) the long-range nature of the interactions is necessary for understanding
the low-temperature properties of spin ice materials.
%
%

\section{\label{sec: spin entropy effects}
Entropic charge: role of the underlying spins
        }
In disregarding the underlying spins in the Debye-H\"{u}ckel approximation to
the free energy of spin ice, we fail to account for quadrupolar corrections to
the monopole description~\cite{Castelnovo2008} (of which we have seen an effect
in the value of the bare monopole cost $\Delta$). We also neglect additional
spin entropic contributions (other than the entropy of mixing of the
monopoles)~\cite{Huse2003,Moessner2003b,Isakov2004,Hermele2004,Henley2005}.

The latter take the form of an entropic charge that adds onto the real magnetic
charge (or, rather, magnetic and entropic coupling constants add) 
for the  monopole Coulomb interactions.
In Appendix~\ref{app: entropic charge} we derive an analytical expression for
the entropic interaction strength and confirm the result by comparing it
to Monte Carlo simulations.
One can then repeat the DH calculations including the entropic correction.
The results are shown in Fig.~\ref{fig: entropic charge effects} (dashed
cyan lines), in comparison to the previous results (solid blue lines),
for the parameters in Sec.~\ref{sec: simulations} with $\DeltaMC = 4.7$~K.
\begin{figure}[ht]
\vspace{0.2 cm}
\begin{center}
\includegraphics[width=0.99\columnwidth]
                {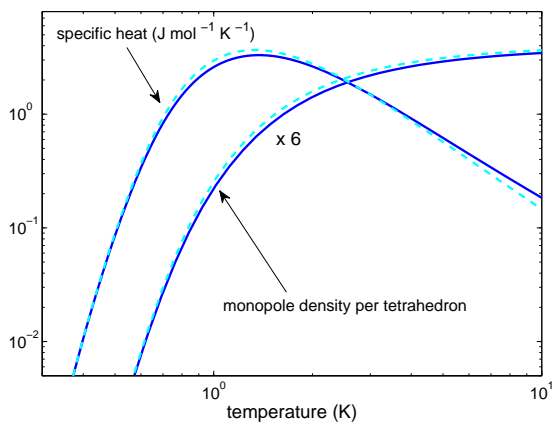}
\end{center}
\caption{
\label{fig: entropic charge effects}
Effects of the entropic charge (dashed cyan lines) on the Debye-H\"{u}ckel
estimate of the heat capacity and monopole density (solid blue lines).
}
\end{figure}
The behaviour of the monopole density and of the heat capacity clearly show
that the entropic contribution can be safely neglected in the low temperature
regime where the DH approximation is valid. It is worth noting that the
relative strength of magnetic and entropic charges can in principle be tuned
straightforwardly, e.g.\ by decreasing $D$ at fixed $J_{\rm eff}$, as the
magnetic monopole charge is proportional to $D$, whereas the scale determining
the applicability of the monopole picture is set by $J_{\rm eff}$.

Indeed, for the nearest-neighbour model with $D=0$, where there is no 
magnetic monopole charge, one would be considering a 
Coulomb gas with entropic interactions only. Debye screening in such a setting 
has already been considered in two dimensions, 
for the  entropic Coulomb gas  encountered in the square lattice 
monomer-dimer model.\cite{krauth2003}
%
%

\section{\label{sec: Comparison to experiment}
Experiment
        }
We now proceed to compare the DH results with experimental data on
{\DTO}. We find good agreement, which is further improved if we use the
latest material parameters from Ref.~\onlinecite{Yavorskii2008}
instead of those in Ref.~\onlinecite{Melko2004}.
Namely, the magnetic moment of the rare earth ions is $9.87$~$\mu_B$
instead of $10$~$\mu_B$; the diamond lattice constant is $4.38$~\AA~
instead of $4.34$~\AA; and the nearest-neighbour exchange coupling
varies between $-3.53$ and $-3.26$, instead of $J = -3.72$~K.

These values result in a new magnetic monopole charge of $4.5$~$\mu_B$/\AA;
a nearest-neighbour interaction strength between monopoles
$E_{\rm nn} = 2.88$~K instead of $3.06$~K;
a dipolar coupling constant $D = 1.32$~K instead of $1.41$~K; and
a bare monopole cost in the range $(4.05,4.23)$~K instead of
$\Delta = 4.35$~K. We reiterate that there are also small corrections due to
further--range superexchange and the quadrupolar interactions, which are not
easily incorporate into the DH framework.
%
%

\subsection{\label{sec: DH secific heat}
Heat capacity
           }
A comparison between the experimentally measured heat capacity and
the one obtained from DH theory,
shows again that the bare monopole cost $\Delta \in (4.05,4.23)$~K
from Eq.~\eqref{eq: Delta} is somewhat too small. Better agreement can be
obtained if, as in the comparison with MC simulations,  we allow for an 8\%
increase in the value of $\Delta \in (4.37,4.57)$~K
(see Fig.~\ref{fig: heat capacity}).
\begin{figure}[ht]
\vspace{0.2 cm}
\begin{center}
\includegraphics[width=0.99\columnwidth]
                {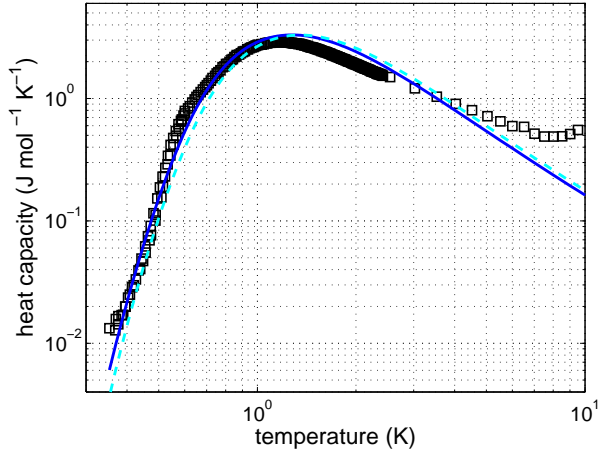}
\end{center}
\caption{
\label{fig: heat capacity}
Experimental results for the heat capacity of {\DTO} (black squares) from
Ref.~\onlinecite{Morris2009}, in units of J/mol~\!K, compared to
the analytical result from Debye-H\"{u}ckel theory with $\Delta = 4.37$~K
(solid blue line) and $\Delta = 4.57$~K (dashed cyan line).
}
\end{figure}
This is in agreement with the results presented in
Ref.~\onlinecite{Morris2009} (Fig.~1), where a value of $\Delta = 4.35$~K\cite{Castelnovo2008}
was used.

\subsection{\label{sec: monopole cost and time scales}
`Dressed' monopole energy and AC susceptibility
           }
The bare monopole cost $\Delta$ is half the energy required for creating and
separating to infinity a pair of monopoles against their long-range Coulomb
attraction. When other monopoles are present, screening effectively truncates
the range of the interactions and there is no further energy cost to
separating a pair beyond the screening length.
In this case it is more appropriate to consider the `dressed' monopole
energy $\Delta_d$ as the energy per monopole that it takes to create a pair
and separate it beyond the screening length. It is indeed the energy
$\Delta_d$ -- rather than $\Delta$ -- that controls for instance the
equilibrium density of the monopoles $\rho \sim e^{-\Delta_d / T}$ at 
intermediate temperatures.

Given the creation energy for a nearest neighbour pair
$\Delta_s = 2\Delta-E_{\rm nn}$ and the expression for the DH screening
length, Eq.~\eqref{eq: DH screening length}, one obtains
\bea
2\Delta_d(T)
&=&
2\Delta - E_{\rm nn}
+
\left( E_{\rm nn} - \frac{\mu_0}{4\pi k_B}\frac{q^2}{\xi_{\rm Debye}(T)} \right)
\nonumber \\
&=&
2\Delta - E_{\rm nn} \frac{a_d}{\xi_{\rm Debye}(T)}
,
\eea
whose behaviour is illustrated in the inset of
Fig.~\ref{fig: dressed monopole energy}.
\begin{figure}[ht]
\vspace{0.2 cm}
\begin{center}
\includegraphics[width=0.99\columnwidth]
                {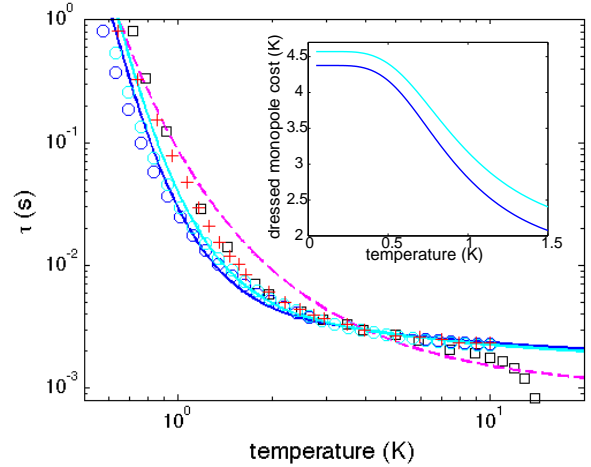}
\end{center}
\caption{
\label{fig: dressed monopole energy}
Experimental magnetic relaxation time scale $\tau$ as a function of temperature
from susceptibility data, Ref.~\onlinecite{Snyder2004} (black open squares).
The rapid increase in $\tau$ at low temperatures is due to the paucity of
defects responsible for the magnetic rearrangement of a spin ice configuration
(namely, the monopoles).
This increase cannot be described by a single exponential (activated
behaviour), as it is evident for instance by comparison with the curve
$\tau = \tau_0\,\exp(\Delta/T)$ (dashed magenta line), say with
$\Delta = 4.5$~K.
On the contrary, a much better agreement is obtained if we replace the bare
monopole energy $\Delta$ with the `dressed' energy $\Delta_d(T)$ (solid blue
curve for $\Delta = 4.37$~K and solid cyan curve for $\Delta = 4.57$~K).
This is compared to $\tau \propto 1/\rho$, where $\rho$ is obtained
from the DH approximation (blue open circles for $\Delta = 4.37$~K and
cyan open circles for $\Delta = 4.57$~K), showing that indeed the
dressing of $\Delta$ accounts for the leading non-exponential correction
in the temperature dependence in the monopole density.
The microscopic time scale was set by imposing that the analytical
results pass through the experimental data point at $4$~K
(see Ref.~\onlinecite{Jaubert2009}).
The inset shows the `dressed' monopole energy $\Delta_d$ as a function of
temperature (solid blue curve for $\Delta = 4.37$~K and solid cyan curve
for $\Delta = 4.57$~K).
}
\end{figure}

A place where this screening effect of the magnetic monopoles becomes
particularly evident is in susceptibility measurements of magnetic relaxation
time scales~\cite{Snyder2004,Jaubert2009}.
Given that the monopoles are responsible for any changes in magnetisation in a
spin ice configuration, the ability of the system to respond to an applied
magnetic field is affected by the monopole density.
For non-interacting monopoles, Ryzhkin showed that in the low temperature, 
hydrodynamic regime the characteristic susceptibility time scale $\tau$ is 
inversely proportional to the monopole density~\cite{Ryzhkin2005}, 
\bea
\tau^{-1} \propto \nu \, T \rho(T)
,
\label{eq: ryzhkin tau}
\eea
where $\nu$ is the mobility of the monopoles. 
This result is likely to be asymptotically correct as $T \to 0$ at zero 
wavevector even in presence of Coulomb interactions, although it is modified 
at finite wavevectors. 

In App.~\ref{app: monopole mobility}, we show that $\nu \sim 1/T$ under
the assumption that Metropolis dynamics are a good approximation to the
microscopic spin flip processes in spin ice.
Therefore,
\bea
\tau \propto 1 / \rho(T)
.
\label{eq: tau = 1_rho}
\eea
As we argued above, at intermediate temperatures $\rho(T)$ is controlled by 
the dressed monopole energy $\Delta_d(T)$ rather than the bare energy 
$\Delta$. Indeed, $\tau$ is poorly fitted by a single
exponential~\cite{Snyder2004,Jaubert2009} such as
$\tau = \tau_0\,\exp(\Delta/T)$.
On the contrary, the curve $\tau = \tau_0\,\exp[\Delta_d(T)/T]$, captures
correctly the faster-than-exponential grows of $\tau$ at low temperatures,
despite the fact that it still significantly underestimates the experimental
value of $\tau$
(see Fig.~\ref{fig: dressed monopole energy}).~\cite{footnote:tau}

Given the good agreement between DH theory and experiments regarding the heat
capacity of the system (Fig.~\ref{fig: heat capacity}) and given that
a similarly good agreement in the heat capacity from Monte Carlo simulations
implied a good agreement also for the monopole density
(Fig.~\ref{fig: monopole density} and Fig.~\ref{fig: MC spec heat}),
one would expect that $\rho(T)$ from Debye-H\"{u}ckel used in
Fig.~\ref{fig: dressed monopole energy} is in fact a good estimate of the
experimental monopole density. 
Therefore, the fact that Eq.~\eqref{eq: tau = 1_rho} underestimates the 
experimental results even when using $\rho(T)$ from DH theory 
is likely due to corrections to the dependence 
$\tau \propto 1 / \rho(T)$ arising from Coulomb interactions at 
intermediate monopole densities. 

At the lowest temperatures (provided of course no ordering or freezing 
intervenes, as it likely would), when monopole separation and screening 
length both diverge, the effective $\Delta_d\rightarrow\Delta$, and hence 
we expect the superexponential behaviour to go away and the curve to follow 
the standard Arrhenius behaviour $\tau \sim \exp(\Delta/T)$. 

From a purely phenomenological perspective, it is interesting to notice 
that a very good agreement beween DH theory and experiments on the
susceptibility time scale $\tau$ (at intermediate temperatures) 
can be obtained by substituting
Eq.~\eqref{eq: tau = 1_rho} with $\tau \propto 1/\rho^{\eta}(T)$,
with $\eta = 3/2$ for $\Delta = 4.37$~K and $\eta = 4/3$ for
$\Delta = 4.57$~K (see Fig.~\ref{fig: tau collapse}).
\begin{figure}[ht]
\vspace{0.2 cm}
\begin{center}
\includegraphics[width=0.99\columnwidth]
                {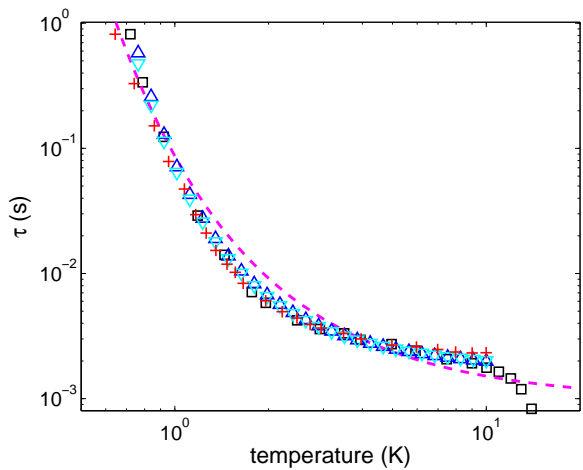}
\end{center}
\caption{
\label{fig: tau collapse}
Experimental magnetic relaxation time scale $\tau$ as a function of 
temperature from susceptibility data, Ref.~\onlinecite{Snyder2004} 
(black open squares). 
The temperature dependence is captures very accurately by a 
phenomenological equation of the type $\tau \propto 1/\rho^\eta$, 
where $\rho$ is obtained from the DH approximation 
(blue upward triangles for $\Delta = 4.37$~K and $\eta = 3/2$; 
cyan downward triangles for $\Delta = 4.57$~K and $\eta = 4/3$). 
The dashed magenta line illustrates the curve 
$\tau = \tau_0\,\exp(\Delta/T)$ with $\Delta = 4.5$~K, for comparison.
}
\end{figure}
Further work is needed to understand the reasons behind such a good overlap.
%
%

\section{\label{sec: beyond DH}
Beyond Debye-H\"{u}ckel
        }
Debye-H\"{u}ckel theory is probably the simplest approximation to obtain the
free energy of a gas of Coulomb interacting particles short of ignoring
interactions altogether.
A number of improvements are available in the vast literature on the
subject~\cite{Fisher1993}, which one can use to obtain a more accurate
description of the magnetic monopole behaviour in spin ice.

Without actually implementing them, we briefly recall hereafter two
common extensions of the DH model.  Firstly, Debye-H\"{u}ckel theory
neglects the association of monopoles into neutral dipolar pairs,
which we have already briefly discussed above (see
Ref.~\onlinecite{Fisher1993} and references therein).  Following
Bjerrum~\cite{Bjerrum1926} (Bj) one can account for such bound pairs,
thus compensating in good part for the uncontrolled linearisation of
the Poisson-Boltzmann equation that is at the basis of the DH
self-consistent solution.  However, whilst being an overall refinement
of DH, DHBj theory leads to unrealistic features in the phase diagram
of the system~\cite{Fisher1993}, with an exponential increase in the
low-temperature fraction of neutral pairs draining the free monopole
density to zero.  This can (and ought to) be compensated by a further
extension to include interactions between dipolar bound pairs and free
monopoles, leading to the so called dipole-ionic (DI)
contribution~\cite{Fisher1993}.  The full DHBjDI theory indeed cures
the unphysical features identified for DHBj, while remaining of course
only an approximation to the exact free energy of the system.

Further improvements on the DHBjDI theory include accounting for
hard-core (HC) effects~\cite{Fisher1993}. It is certainly worthwhile
developing the theory further in this direction, especially in
settings or for quantitites where new phenomena (e.g., a dominant
population of bound pairs), rather than only quantitative corrections,
ensue.
%
%

\section{\label{sec: conclusions}
Conclusions
        }
In summary, we have presented a theory for the low-temperature physics of 
spin ice within the Debye-H\"uckel framework familiar from the study of 
(electric) Coulomb liquids. The success of this simple approach in treating 
the low-energy physics of spin ice is a testament to the power of the 
`variable transformation' from magnetic dipoles to magnetic monopoles 
appropriate to the Coulomb phase with its emergent gauge field. 

With this first step accomplished, next on the wishlist are a number of 
items some of which should push our attention beyond the framework provdided 
by the DH paradigm. Firstly, a more detailed understanding of spin ice 
(hydro-)dynamics; secondly, an extension of this theory to a broader class 
of parent Hamiltonians, perhaps even including coherent quantum dynamics; 
and thirdly, contact with all the non-equilibrium experiments suggesting 
that not only the sparseness of monopoles but also phononic physics plays 
a role in the freezing of spin ice around 
$T_f$.~\cite{Slobinsky2010}
%
%

\section*{
Acknowledgments
         }
We are very grateful to our experimental collaborators of 
Ref.~\onlinecite{Morris2009} -- in particular Santiago Grigera, Klaus Kiefer, 
Bastian Klemke, Michael Meissner, Jonathan Morris, Kirilly Rule, 
Damian Slobinsky and Alan Tennant -- for the discussions and experimental 
measurements which motivated us to pursue the Coulomb gas analogy
in detail as reported here.

This work was supported in part by EPSRC Postdoctoral Research Fellowship
EP/G049394/1 (C.C.) and by NSF Grant Number DMR-1006608 (S.L.S.). 
We mutually acknowledge hospitality and travel support 
for visits to our respective institutions. 
%
%

\appendix

\section{\label{app: single tet approx}
Single tetrahedron approximation
        }
An alternative approximation that can be used to obtain the spin ice free
energy and related thermodynamic quantities is to use a truncated cluster
expansion. Most simply, this amounts to computing explicitly the free
energy of an isolated tetrahedron by direct summation over all $2^4$ states.

At this level, all interactions are
nearest-neighbour ones. In terms of this effective short range coupling
$J_{\rm eff}$, the partition function of a tetrahedron is
\bea
Z
&=&
\left[
  6 + 8 e^{-2 J_{\rm eff}/T} + 2 e^{-8 J_{\rm eff}/T}
\right]^{N_t}
.
\eea
From this, one can estimate the partition function of the entire system,
\bea
Z
&=&
2^{N_s}
\left[
  \frac{6 + 8 e^{-2 J_{\rm eff}/T} + 2 e^{-8 J_{\rm eff}/T}}{16}
\right]^{N_s/2}
,
\eea
and thus the free energy per spin in degrees Kelvin,
$F/N_s k_B = - (T/N_s) \ln Z$.

Substituting into Eq.~(\ref{eq: spec heat from free energy}),
%
%
we obtain the heat capacity of the system (in units of J/K per Dy ion),
\beq
c_V
=
\frac{24 k_B J_{\rm eff}^2}{T^2}
\frac{e^{6 J_{\rm eff} / T}
      \left( 3 - 2 e^{2 J_{\rm eff} / T} + e^{4 J_{\rm eff} / T}\right)}
     {\left(
        1
        - e^{2 J_{\rm eff} / T}
        + e^{4 J_{\rm eff} / T}
        + 3 e^{6 J_{\rm eff} / T}
      \right)^2}
.
\eeq

The choice of $J_{\rm eff} = 5D/3 + J/3 = 1.11$~K, which corresponds to 
the nearest-neighbour interaction strength from the exchange plus dipolar 
coupling constants, yields a very poor agreement with the experimental data 
(not shown). The situation improves slightly if we take advantage of the 
projective equivalence between dipolar and nearest-neighbour interactions on 
the pyrochlore lattice~\cite{Isakov2005}. 
Instead of truncating the dipolar contribution to $5D/3$, one can therefore
use the effective value of $J_{\rm nn}$ that yields the same low-energy
spectrum as from the long range dipolar interactions.
This value can be derived using the dumbell decomposition in
Ref.~\onlinecite{Castelnovo2008}, $J_{\rm eff} = 1.45$~K.
\iffalse
The result is shown in Fig.~\ref{fig: heat capacity} (dash dotted yellow line)
and it is indeed in quantitative agreement with the experimental data at high
temperatures $T \gtrsim 2$~K, as expected of a cluster expansion of the free
energy.
\else
The result is shown in Fig.~1 of Ref.~\onlinecite{Morris2009}
and it is indeed in quantitative agreement with the experimental data at high
temperatures $T \gtrsim 2$~K, as expected of a cluster expansion of the free
energy.
\fi

Note that even if we allow $J_{\rm eff}$ to vary as a fitting parameter in
the theory, the shape of $c_V(T)$ does not change significantly and it can
be brought to agree with the experimental data only over a very narrow
temperature interval.
By comparison, this highlights even more how effective the Debye-H\"{u}ckel
free energy is at capturing the low energy fluctuations in dipolar spin
ice.
%
%

\section{\label{app: entropic charge}
Entropic monopole charge
        }
The effective description of spin-ice in the absence of monopoles is given
by the probability distribution of a magnetostatic-like (divergenceless)
field~\cite{Henley_coulomb}
\bea
\mathcal{P}
\propto
\exp\left[
-\frac{\mathcal{K}}{2} v^{-1}_{\rm cell}
  \int \left\vert \vec{B}^{\rm ent}(r) \right\vert^2 \:d^3r
\right]
\label{eq: entropic term}
\\
\times
\exp\left[
-\frac{\mu_0}{2 k_B T}
  \int \left\vert \vec{H}^{\rm mag}(r) \right\vert^2 \:d^3r
\right]
.
\label{eq: magnetic term}
\eea

The first term Eq.~\eqref{eq: entropic term} is purely entropic in origin.
The geometric field
$\vec{B}^{\rm ent}(r)$ is obtained from coarse graining fixed-length
vectors that identify the local direction of the spins in the system.
Here $v_{\rm cell}$ is the volume of the primitive unit cell 
(Fig.~\ref{fig: lattice conventions}).
Introducing the coarse grained (dimensionless) field
$\vec{B}(r)$ defined at the centre of each tetrahedron (belonging to one of
the two sublattices) as
\bea
\left(
\begin{array}{c}
B_x\\
B_y\\
B_z\\
\end{array}
\right)
&=&
\frac{1}{\sqrt{3}}
\left(
\begin{array}{cccc}
1 & 1 & -1 & -1\\
1 & -1 & 1 & -1\\
1 & -1 & -1 & 1\\
\end{array}
\right)
\left(
\begin{array}{c}
S_0\\
S_1\\
S_2\\
S_3\\
\end{array}
\right)
\label{eq: coarse graining}
\eea
the stiffness coefficient can be determined to be $\mathcal{K} = 3/8$.
(Note that we used a different field normalisation with respect to
Ref.~\onlinecite{Conlon_ref}, so as to preserve the underlying spin length
equal to $1$.)
\begin{figure}[ht]
\vspace{0.2 cm}
\begin{center}
\includegraphics[width=0.95\columnwidth]
                {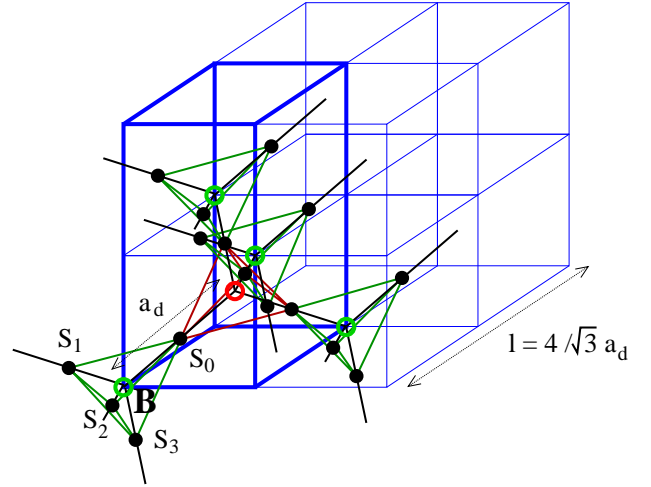}
\end{center}
\caption{
\label{fig: lattice conventions}
Lattice conventions.
The highlighted portion of the blue cube (i.e., the 16-spin cubic unit cell
in spin ice) corresponds to a possible choice of the primitive unit cell
in the fcc lattice formed by the centres of one sublattice of tetrahedra
in the pyrochlore lattice (circled in green in the figure).
}
\end{figure}

The second term Eq.~\eqref{eq: magnetic term} accounts for the magnetic
energy stored in a spin ice configuration (devoid of monopoles).
In this case, $\vec{H}^{\rm mag}(r)$ is the magnetic field generated
by the spin magnetic moments $\mu$ pointing in the local spin direction
($\mu_0$ is the permeability of the vacuum, $k_B$ is the Boltzmann constant
and $T$ is the temperature of the system).

Given that the total field $\vec{B} = \mu_0 (H + M)$ is always divergenceless,
the field $\vec{H}^{\rm mag}(r)$ can be equivalently replaced by the
magnetisation per unit volume $M$, which in turn can be obtained
by coarse graining the spin magnetic moments.
Using the scheme~\eqref{eq: coarse graining} already adopted for
$\vec{B}^{\rm ent}(r)$ over a primitive unit cell, we have that
\bea
\left\vert \vec{H}^{\rm mag}(r) \right\vert
=
\left\vert \vec{M}(r) \right\vert
=
\frac{\mu}{v_{\rm cell}} \left\vert \vec{B}^{\rm ent}(r) \right\vert
.
\eea
Therefore, the difference between the two terms Eq.~\eqref{eq: entropic term}
and Eq.~\eqref{eq: magnetic term} can be reduced to different coefficients
\bea
\frac{\mathcal{K}}{v_{\rm cell}}
\quad
{\rm vs}
\quad
\frac{\mu_0 \mu^2}{k_B T v^2_{\rm cell}}
\label{eq: coefficients}
\eea
to the same integral
$\int \vert \vec{B}^{\rm ent}(r) \vert^2 \:d^3r$.

It is convenient to re-express the magnetic coefficient in terms of the
magnetic Coulomb energy of two monopoles placed in adjacent tetrahedra
(expressed in degrees Kelvin),
\bea
E_{\rm nn}
=
\frac{\mu_0}{4\pi k_B}\frac{q^2}{a_d}
=
\frac{\mu_0}{\pi k_B}\frac{\mu^2}{a^3_d}
\\
\Rightarrow
\;\;\;
\frac{\mu_0 \mu^2}{k_B T v^2_{\rm cell}}
=
\frac{E_{\rm nn}}{T}\frac{\pi a^3_d}{v^2_{\rm cell}}
,
\eea
where we used the fact that $q=2\mu/a_d$, $a_d$ being the diamond lattice
constant.
By comparison with the entropic coefficient, we can then identify the
entropic counterpart to the neareast-neighbour Coulomb energy,
\bea
\frac{E^{\rm ent}_{\rm nn}}{T}\frac{\pi a^3_d}{v^2_{\rm cell}}
=
\frac{\mathcal{K}}{v_{\rm cell}}
\\
\Rightarrow
\;\;\;
\frac{E^{\rm ent}_{\rm nn}}{T}
=
\frac{\mathcal{K}}{\pi} \frac{v_{\rm cell}}{a^3_d}
.
\eea

If we finally use the fact that $v_{\rm cell}$ is $1/4$ of the volume of
the 16-spin cubic unit cell in spin ice, $v = (4 a_d / \sqrt{3})^3$,
and that with the coarse graining~\eqref{eq: coarse graining}
$\mathcal{K} = 3/8$, we arrive at the result
\bea
\frac{E^{\rm ent}_{\rm nn}}{T}
=
\frac{\mathcal{K}}{\pi} \frac{16}{3 \sqrt{3}}
=
\frac{2}{\sqrt{3} \pi}
\simeq
0.36755
.
\label{eq: E_nn entropic analytical}
\eea

It is interesting to convert this value into an entropic monopole charge :
\bea
q_{\rm ent}
&=&
\sqrt{\frac{4\pi a_d k_B E^{\rm ent}_{\rm nn}}{\mu_0}}
=
1.48\:10^{-13}\:\sqrt{T}
\nonumber \\
&=&
1.6 \: \sqrt{T} \; \mu_B / \textrm{\AA}
.
\eea
The entropic charge of a monopole becomes larger than the
real magnetic charge only for $T \gtrsim 8$~K, well beyond the limit of
validity of the monopole description of spin ice.
In the experimentally relevant temperature range $0.1-1$~K, the entropic
contribution ranges from $1$\% to $10$\% of the real magnetic contribution
to the energy of the monopoles.

In order to confirm this analytical estimate of the entropic Coulomb
interaction strength in spin ice, we have run Monte Carlo simulations of
the nearest-neighbour spin ice model, sampling only
configurations with two monopoles (one positive, one negative).
Such configurations are all isoenergetic and the monopole positions can be
updated at every Monte Carlo step without rejection.
Ergodicity was tested by computing spin-spin autocorrelation
functions. The distribution of separation distances between the two monopoles
was then sampled both in Monte Carlo time and across different initial
configurations and random number seeds.
%
%
\begin{figure}[ht]
\vspace{0.2 cm}
\begin{center}
\includegraphics[width=0.95\columnwidth]
                {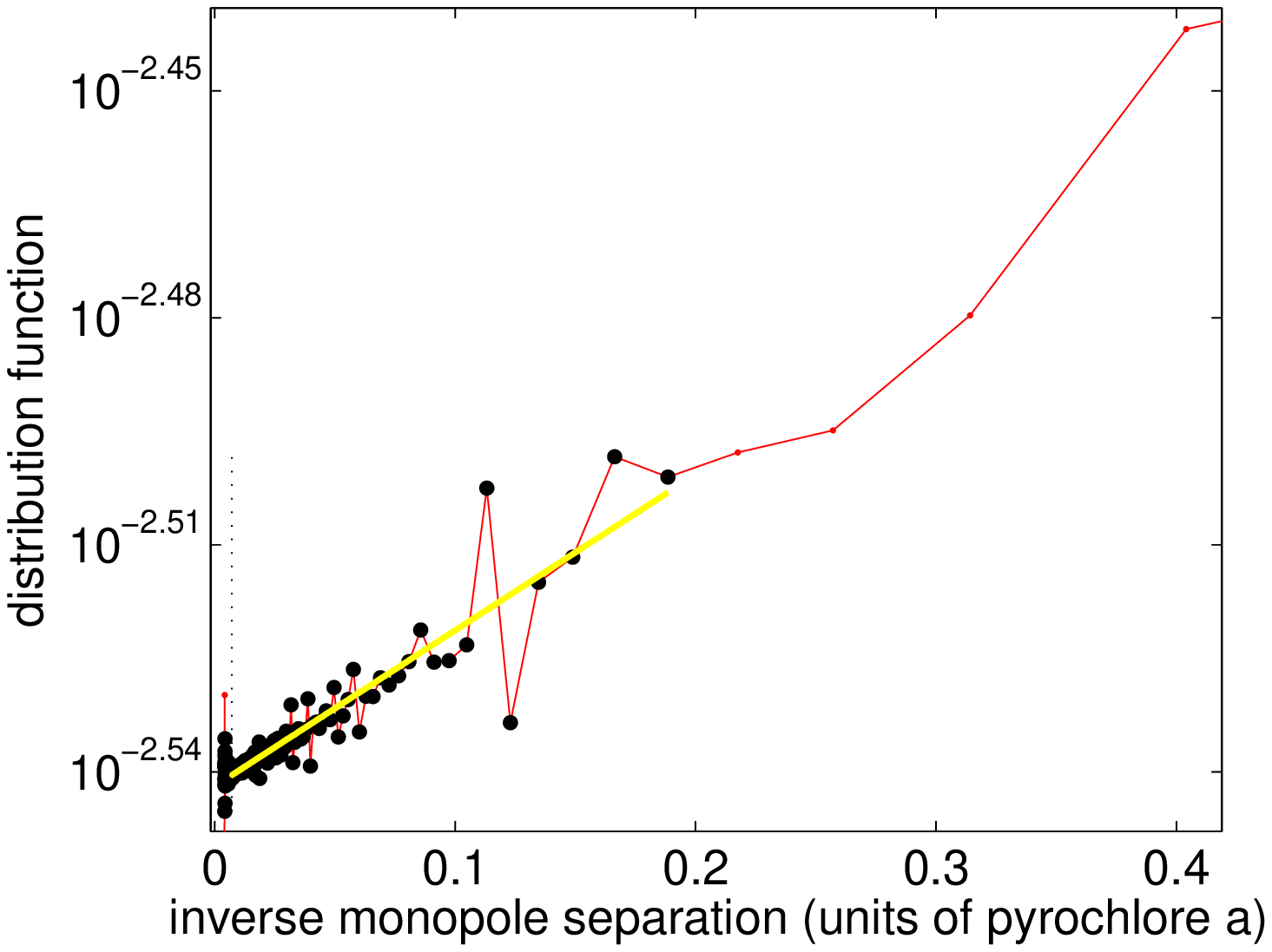}
\\
\includegraphics[width=0.95\columnwidth]
                {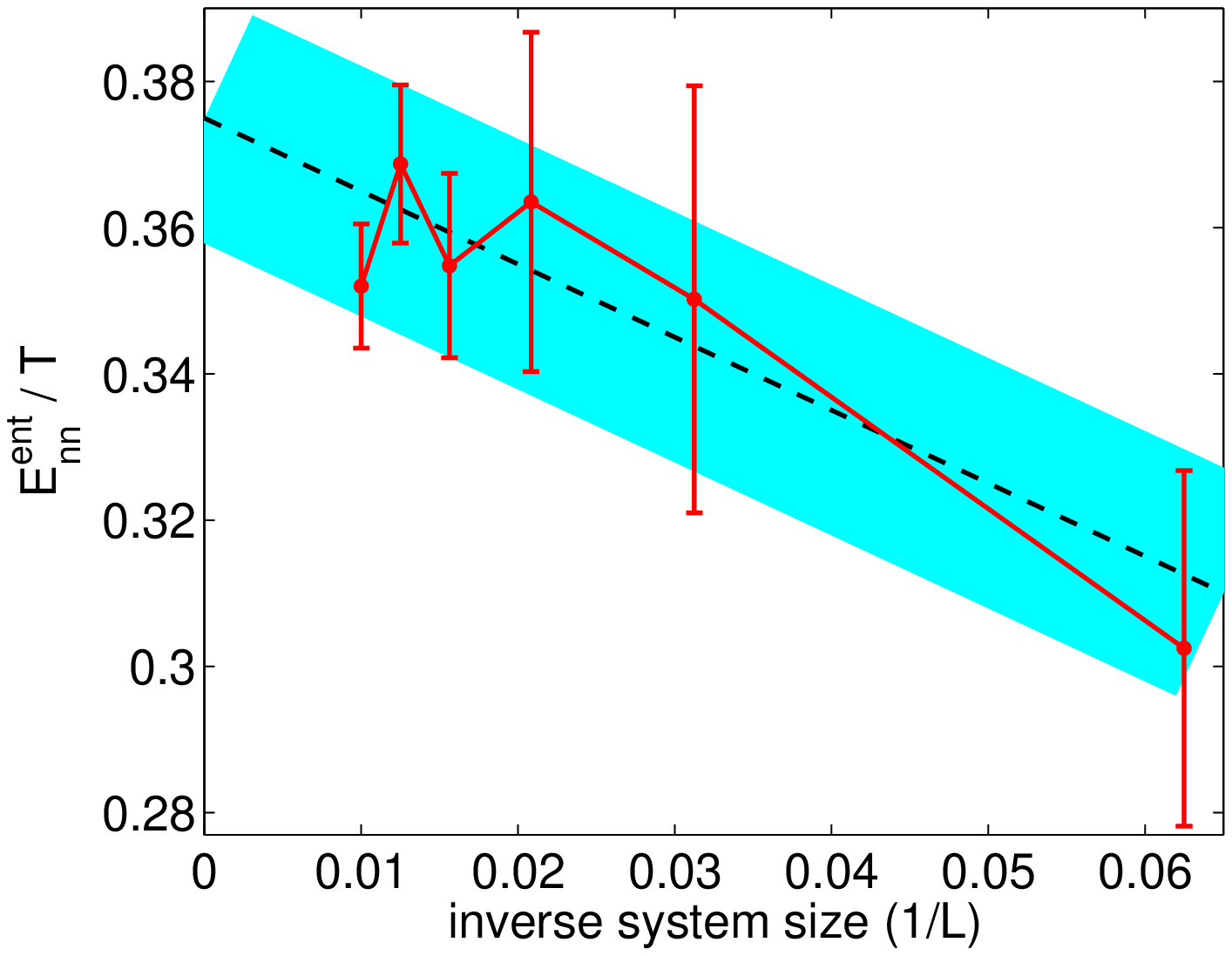}
\end{center}
\caption{
\label{fig: entropic charge fit}
Top Panel:
Distribution of distances per lattice site between two monopoles
in a spin ice configuration of $16 \times L^3$ spins, $L=64$.
(top panel, red curve).
The expected form due to the entropic Coulombic interaction is
$\mathcal{P}\sim \exp(E_{\rm nn}^{\rm ent}/T R)$ and the solid yellow
line is the linear fit of $\ln P(R)$ as a function of $1/R$.
Bottom Panel:
Finite size scaling of the nearest neighbour entropic interaction
$E_{\rm nn}^{\rm ent} / T$ vs. the inverse system size $1/L$,
$L=16,\,32,\,48,\,64,\,80,\,100$.
The dashed black line and shaded cyan region are a guide to the eye for
a reasonable $L \to \infty$ extrapolation and confidence interval,
leading to $E_{\rm nn}^{\rm ent}/T \simeq 0.375 \pm 0.015$.
}
\end{figure}

{}From Eq.~\ref{eq: entropic term}, it follows that the entropic interaction
between the two monopoles leads to a probability distribution of the form
$\mathcal{P}(R) \sim R^2 \, \exp(E_{\rm nn}^{\rm ent}/T R)$,
where $R$ is the separation distance in units of the diamond lattice spacing.
In particular, if we sample the distribution \emph{per lattice site} at
distance $R$, it has a purely exponential form
$\sim \exp(E_{\rm nn}^{\rm ent}/T R)$,
and one can obtain the value of $E_{\rm nn}^{\rm ent}/T$ from linear
fits in semi-logarithmic scale (Fig.~\ref{fig: entropic charge fit},
top panel).

We repeated these fits for different system sizes in order to account for
finite size scaling (illustrated in Fig.~\ref{fig: entropic charge fit},
bottom panel).
Even though the accuracy of our simulations does not allow for a reliable
extrapolation in the $L\to\infty$ limit, the nearest-neighbour entropic
interaction strength appears to lie in the interval
$E_{\rm nn}^{\rm ent}/T \simeq 0.375 \pm 0.015$,
in reasonable agreement with the analytical value in
Eq.~\eqref{eq: E_nn entropic analytical}, $2/\sqrt{3} \pi \simeq 0.36755$.
%
%

\section{\label{app: monopole mobility}
Monopole mobility
        }
The mobility of the monopoles in spin ice (and thus its temperature
dependence) can be estimated from microscopic considerations, under the
assumption that Metropolis-like equations govern the dynamics of the
system~\cite{Castelnovo2010,Castelnovo2011}.

The mobility of a particle is given by the ratio of its drift velocity $v_d$
over the driving force strength $q E$, $\nu = v_d / (q E)$.

Under Metropolis dynamics for a particle with charge $q$ in
a field $E$, the average displacement in a single step is
\beq
\Delta x
=
\ell \, \frac{1 - e^{-\beta q V}}{1 + e^{-\beta q V}}
,
\eeq
where $\ell$ is the characteristic microscopic length scale,
$V$ is the potential difference for a single hopping process,
$1$ is the probability to hop in the direction of the field, and
$\exp(-\beta q V)$ is the probability to hop in the opposite direction.

Note that, on a lattice, there can be several inequivalent forward and
backward hoppings, depending on the direction of the field.
For example, while a $45^\circ$ field applied to charged particles living on
a square lattice is described straightforwardly by the above equation
(with $\ell = a/ \sqrt{2}$, $a$ being the lattice spacing), a $90^\circ$ field
on the same lattice allows for a forward, a backward, and two perpendicular
hopping processes (see Fig.~\ref{fig: hopping direction}).
\begin{figure}[ht]
\vspace{0.2 cm}
\begin{center}
\includegraphics[width=0.8\columnwidth]
                {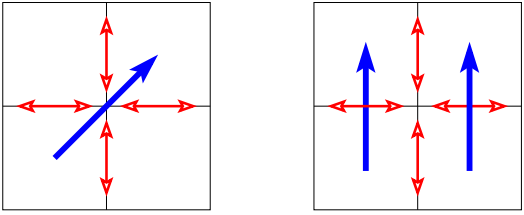}
\end{center}
\caption{
\label{fig: hopping direction}
Two examples of how the available hopping processes depend on the direction
of the applied field on a square lattice: a $45^\circ$ field (left) and
a $90^\circ$ field (right).
}
\end{figure}
One therefore needs to average over all of them to obtain the correct value
of $\Delta x$.

For convenience, we choose to define the mobility $\nu$ as
\bea
\frac{\Delta x/a}{\tau_0}
&=&
\frac{\ell}{a \tau_0} \frac{1 - e^{-\beta q V}}{1 + e^{-\beta q V}}
\\
&\equiv&
\nu \, q E a
,
\eea
for small values of the applied field $E$. Here $a$ is the (dimensionful)
lattice constant and $\tau_0$ is the microscopic time scale for a single
MC step.
At large temperatures with respect to the field strength, one can expand the
exponentials and arrive at the expression
\bea
\nu
&=&
\frac{1}{\tau_0}
\frac{1}{q E a}
\frac{\ell}{a}
\frac{1 - e^{-\beta q V}}{1 + e^{-\beta q V}}
\\
&=&
\frac{1}{\tau_0}
\frac{\ell}{a} \frac{V / (E a)}{2 k_B T}
+
\mathcal{O}\left[ \frac{(\beta V)^2}{q E a} \right]
.
\label{eq: MC mobility}
\eea
%
%

For example, the case of a generic field direction on the anisotropic square
lattice, with lattice constants $a$ and $b$, gives
\begin{widetext}
\bea
\nu
&=&
\frac{1}{\tau_0}
\frac{1}{q E a^2}
\frac{a\cos\theta + b\sin\theta
      - a\cos\theta e^{-\beta q E a\cos\theta}
      - b\sin\theta e^{-\beta q E b\sin\theta}}
     {1 + 1
      + e^{-\beta q E a\cos\theta}
      + e^{-\beta q E b\sin\theta}}
\nonumber \\
&\simeq&
\frac{1}{\tau_0}
\frac{1}{4 k_B T}
\frac{a^2\cos^2\theta + b^2\sin^2\theta}{a^2}
+
\mathcal{O}\left( \beta^2 E \right)
.
\eea
\end{widetext}
If the lattice is isotropic ($a=b$) the mobility is independent
of the direction of the applied field,
\bea
\nu
&\simeq&
\frac{1}{\tau_0}
\frac{1}{4 k_B T}
+
\mathcal{O}\left( \beta^2 E \right)
.
\eea

The mobility of monopoles on an isotropic diamond lattice, of lattice
constant $a_d$, with respect to a generic field direction $\hat{e}$ can
be computed in a similar way, with the additional care that there are now two
inequivalent sublattices.
With respect to one sublatice, we obtain
\begin{widetext}
\bea
\nu
&=&
\frac{1}{\tau_0}
\frac{1}{q E a_d}\frac{1}{\sqrt{3}}
\frac{(\hat{e}_1+\hat{e}_2+\hat{e}_3) \:
      \min\left[
        1, e^{\beta q E a_d (\hat{e}_1+\hat{e}_2+\hat{e}_3)/\sqrt{3}}
      \right]
      +
      (\hat{e}_1-\hat{e}_2-\hat{e}_3) \:
      \min\left[
        1, e^{\beta q E a_d (\hat{e}_1-\hat{e}_2-\hat{e}_3)/\sqrt{3}}
      \right]}
     {\min\left[
        1, e^{\beta q E a_d (\hat{e}_1+\hat{e}_2+\hat{e}_3)/\sqrt{3}}
      \right]
      +
      \min\left[
        1, e^{\beta q E a_d (\hat{e}_1-\hat{e}_2-\hat{e}_3)/\sqrt{3}}
      \right]}
\nonumber \\
&&
\frac{+
      (-\hat{e}_1+\hat{e}_2-\hat{e}_3) \:
      \min\left[
        1, e^{\beta q E a_d (-\hat{e}_1+\hat{e}_2-\hat{e}_3)/\sqrt{3}}
      \right]
      +
      (-\hat{e}_1-\hat{e}_2+\hat{e}_3) \:
      \min\left[
        1, e^{\beta q E a_d (-\hat{e}_1-\hat{e}_2+\hat{e}_3)/\sqrt{3}}
      \right]}
     {+
      \min\left[
        1, e^{\beta q E a_d (-\hat{e}_1+\hat{e}_2-\hat{e}_3)/\sqrt{3}}
      \right]
      +
      \min\left[
        1, e^{\beta q E a_d (-\hat{e}_1-\hat{e}_2+\hat{e}_3)/\sqrt{3}}
      \right]}
\nonumber \\
&\simeq&
\frac{1}{\tau_0}
\frac{1}{12 k_B T}
\left[
  (\hat{e}_1+\hat{e}_2+\hat{e}_3)^2 \,
  \Theta_<(\hat{e}_1+\hat{e}_2+\hat{e}_3)
  +
  (\hat{e}_1-\hat{e}_2-\hat{e}_3)^2 \,
  \Theta_<(\hat{e}_1-\hat{e}_2-\hat{e}_3)
\right.
\nonumber \\
&&
\qquad\;\;
\left.
  +
  (-\hat{e}_1+\hat{e}_2-\hat{e}_3)^2 \,
  \Theta_<(-\hat{e}_1+\hat{e}_2-\hat{e}_3)
  +
  (-\hat{e}_1-\hat{e}_2+\hat{e}_3)^2 \,
  \Theta_<(-\hat{e}_1-\hat{e}_2+\hat{e}_3)
\right]
+
\mathcal{O}\left( \beta^2 E \right)
,
\label{eq: mobility diamond sub 1}
\eea
where $\Theta_<(x)=\Theta(-x)$ is the Heaviside theta function.
With respect to the other sublatice, we obtain
\bea
\nu
&\simeq&
\frac{1}{\tau_0}
\frac{1}{12 k_B T}
\left[
  (\hat{e}_1+\hat{e}_2+\hat{e}_3)^2 \,
  \left[ 1-\Theta_<(\hat{e}_1+\hat{e}_2+\hat{e}_3) \right]
  +
  (\hat{e}_1-\hat{e}_2-\hat{e}_3)^2 \,
  \left[ 1-\Theta_<(\hat{e}_1-\hat{e}_2-\hat{e}_3) \right]
\right.
\nonumber \\
&&
\qquad\;\;
\left.
  +
  (-\hat{e}_1+\hat{e}_2-\hat{e}_3)^2 \,
  \left[ 1-\Theta_<(-\hat{e}_1+\hat{e}_2-\hat{e}_3) \right]
  +
  (-\hat{e}_1-\hat{e}_2+\hat{e}_3)^2 \,
  \left[ 1-\Theta_<(-\hat{e}_1-\hat{e}_2+\hat{e}_3) \right]
\right]
+
\mathcal{O}\left( \beta^2 E \right)
.
\nonumber \\
\label{eq: mobility diamond sub 2}
\eea
\end{widetext}
If we finally take the average of both sublattices, we arrive at
\bea
\nu
&\simeq&
\frac{1}{2 \tau_0}
\frac{1}{12 k_B T}
\left[
  (\hat{e}_1+\hat{e}_2+\hat{e}_3)^2
  +
  (\hat{e}_1-\hat{e}_2-\hat{e}_3)^2
\right.
\nonumber\\
&&\qquad\qquad\:\:
\left.
  +
  (-\hat{e}_1+\hat{e}_2-\hat{e}_3)^2
  +
  (-\hat{e}_1-\hat{e}_2+\hat{e}_3)^2
\right]
\nonumber\\
&+&
\mathcal{O}\left( \beta^2 E \right)
\nonumber \\
&\simeq&
\frac{1}{6}
\frac{1}{\tau_0}
\frac{1}{k_B T}
+
\mathcal{O}\left( \beta^2 E \right)
,
\eea
independently of the direction of the field $E$.

If the magnetic monopoles on the diamond lattice are in fact the collective
excitations in a spin ice system, one needs to take into account the
constraint that one of the three possible hopping directions is essentially
forbidden, as it would create doubly charged excitations.
Taking the average over the possible forbidden directions does not introduce
a dependence on the field direction
and we can therefore choose to compute the mobility in a $[100]$ magnetic
field for convenience:
\bea
\nu
&=&
\frac{1}{2\tau_0}
\frac{1}{q E a_d}
\frac{1}{\sqrt{3}}
\frac{2 - e^{-\beta q E (a_d/\sqrt{3})}}
     {2 + e^{-\beta q E (a_d/\sqrt{3})}}
\nonumber\\
&+&
\frac{1}{2\tau_0}
\frac{1}{q E a_d}
\frac{1}{\sqrt{3}}
\frac{1 - 2e^{-\beta q E (a_d/\sqrt{3})}}
     {1 + 2e^{-\beta q E (a_d/\sqrt{3})}}
\nonumber \\
&\simeq&
\frac{4}{27} \frac{1}{\tau_0} \frac{1}{k_B T}
,
\eea

Notice that these results are independent of whether the potential and field 
had an entropic or magnetic origin, provided that the assumption of
the field being smooth over distances of the order of the lattice spacing
$a_d$ holds.
This definition of the mobility shows in fact that it depends only on some
microscopic time scale $\tau_0$ and on the thermal energy per particle in the
system.
%
%


\begin{thebibliography}{99}
\bibitem{footnote:dipolar}
Note that the correlations have a dipolar piece at $T > 0$ also, due to the 
long ranged interactions.
\end{thebibliography}
\end{document}